\documentclass[twocolumn,prl,aps,showpacs,superscriptaddress,floatfix,showkeys,nofootinbib,10pt]{revtex4-1}

\usepackage{amsmath,amsfonts,amssymb}
\usepackage{graphicx}
\usepackage{mathtools}
\usepackage{color}

\newcommand{\ket}[1]{ | \, #1 \rangle} \newcommand{\bra}[1]{ \langle #1 \, |} 
\newcommand{\proj}[1]{\ket{#1}\bra{#1}} 
\newcommand{\kb}[2]{\ket{#1}\bra{#2}}
\newcommand{\quot}[1]{``#1''}
\newcommand{\E}[1]{ \langle #1 \, \rangle} 
\newcommand{\TN}[1]{ \left| \left| #1 \, \right| \right|_1} 
\newcommand{\Ab}[1]{ \left| #1 \, \right|} 
\newcommand{\be}{\begin{equation}} \newcommand{\ee}{\end{equation}}
\newcommand{\ba}{\begin{aligned}} \newcommand{\ea}{\end{aligned}}
\newcommand{\Nm}{N_{\text{mac}}}

\DeclareMathOperator{\Tr}{Tr}

\newtheorem{prop}{Proposition}
\newtheorem{cor}{Corollary}

\begin{document}

\title{Monitoring of the process of system information broadcasting in time}

\author{P.~Mironowicz} \email{piotr.mironowicz@gmail.com} \affiliation{Department of Algorithms and System Modeling, Faculty of Electronics, Telecommunications and Informatics, Gda\'nsk University of Technology} \affiliation{National Quantum Information Centre in Gda\'nsk,
81-824 Sopot, Poland}
\author{J.~K.~Korbicz} \affiliation{Faculty of Applied Physics and Mathematics, Gda\'nsk University of Technology, 80-233 Gda\'nsk, Poland} \affiliation{National Quantum Information Centre in Gda\'nsk, 81-824 Sopot,
Poland}
\author{P.~Horodecki} \email{pawel@mif.pg.gda.pl} \affiliation{Faculty of Applied Physics and Mathematics, Gda\'nsk University of Technology, 80-233 Gda\'nsk, Poland} \affiliation{National Quantum Information Centre in Gda\'nsk, 81-824 Sopot, Poland}

\date{\today}

\begin{abstract}

One of the fundamental problems of modern physics is how the classical world, the 2nd Law of Thermodynamics and the whole irreversibility emerges from the quantum reality with reversible evolution. This relates to the problem of measurement transforming quantum, non-copyable data, towards intersubjective, copyable classical knowledge. We use the quantum state discrimination to show in a central system model how it's evolution leads to the broadcasting of the classical information. We analyze the process of orthogonalization and decoherence, their time scales and dependence on the environment.
	
\end{abstract}


\maketitle

One of the fundamental questions in ontology is: what does it mean to exist?, and the related question: what does exist? The history of different answers to these questions is full of conjectures and refutations. We sketch only a few of them, see \cite{Ingarden} for comprehensive discussion.

In Ancient times Protagoras said that everything that is perceived, exists; from this and the relativity of perception he conjectured that the reality is relative itself. Epicurus assigned the real existence even to dreams, as perceived by some person. They both refuted the principle of non-contradiction. On the other hand, Plato said that only Ideas (Forms) exist, and perceptions are just stimuli for recollection of the knowledge. Aristotle's view, called Hylomorphism, was that the existing objects (substances) are constituted by both unperceived matter and perceivable form. For Plato forms existed on their own, for Aristotle they existed in real things.

Further positions in the debate were stated by G.~Berkeley and I.~Kant. The former said in his famous quote \quot{esse est percipi}, to be is to be perceived (by spirits, i.e. conscious subjects or observers). Since the perception is relative, he claimed that there exist sensations, but there is no real things (in consequence, he concluded that God has to be the source of sensations) \cite{Berkeley}. Inspired by this view, Kant came to different conclusions. He distinguished the unintelligible but existing things-in-itself (noumenon) and things that appear (phenomenon), perceived by a reason via senses of a particular being \cite{Kant} (cf. e.g. \cite{Heller} for essentialism vs. phenomenalism discussion).

Anyhow different are the views stated above, they have in common the intuition that from the (natural) assumption that the existing things cannot be contradictory, it follows that they should not be perceived in different ways by different observers. This idea of objectivity can be formalized in the following way: A state of the system S exists objectively if many observers can find out the state of S independently, and without perturbing it \cite{ZurekNature,myObj}.
We stress that this notion is purely epistemic, and it seems to be weaker than the ontological one, nonetheless it complies with modern approaches to the meaning of the objectivity of the knowledge defying essentialism~\cite{Popper}.
(One may consider an even weaker concept of the objectivity, see \cite{Brandao15}.)

A fundamental step towards understanding the emergence of objectivity from quantum physics has been put forward by W. H. \.Zurek \cite{Zurek81,Zurek82}
and is known as the quantum Darwinism idea \cite{Zurek06,ZurekNature}. This elaborated form of decoherence theory \cite{SchlosshauerRMP} states in its essence that if the same information about a system of interests is being efficiently proliferated into different fractions of the environment, and in consequence redundantly imprinted and stored in them, it becomes objective (or, more precisely, intersubjective in the meaning of K.~Ajdukiewicz \cite{Ajdukiewicz49,AjdukiewiczEng}) and classical \cite{Zurek04}
in the sense that the fractions may be accessed independently by many observers gaining the same or similar knowledge. On the other hand, the no-cloning theorem \cite{nocloning1,nocloning2} forbids a direct copying of a state of the system to different fractions, and even a weaker form of distribution of quantum states, the state broadcasting, is not always possible \cite{broadcast96}. 

It has been shown recently \cite{myObj} that the emergence of the classical and objective properties in the spirit recalled above is due to
a form of information broadcasting (similar to the so called \quot{spectrum broadcasting} \cite{broadcast12}) leading to a creation of a specific quantum state structures between the system and a fraction of its environment called Spectrum Broadcast Structures (SBSs).
It must be stressed that some variant of the emergence of the objectivity in a weaker terms of \quot{objectivity of observables} has been proven to be a universal phenomenon \cite{Brandao15}. However it is SBS 
that ensures a perfect access of different observers to some property of the observed systems. 
However, the SBSs encountered so far \cite{myPRL,TuziemskiKorbiczEPL,TuziemskiKorbiczPhotonics} were reached in the course of the evolution in the asymptotic limit only (even the dynamical SBS of \cite{TuziemskiKorbiczEPL}). What if we would like to know how far is the actual state of a system from being objective? Here we answer this question in a model independent way by providing a natural construction of an ideal SBS, approximating at any given moment the actual state of the system and a fraction of its environment. The construction is based on the state overlaps and decoherence factors, as well as the quantum state discrimination \cite{Helstrom,Qiu}. We consider here the important quantum measurement limit and show how SBS occurs for time scales large enough and for large enough coarse-grainings of the environment, called macrofractions \cite{myPRL}.

\textit{\bf Approximate, momentary SBS via state distinguishability. -} We start with a general situation of a central system~$S$ interacting with $M$-partite environment, with $\varrho(0) = \varrho_{S}(0) \otimes \bigotimes_{k=1}^{M} \varrho^{(k)}(0)$. We assume quantum measurement limit, were the central interaction Hamiltonian $H_{int} = A \otimes \sum_{k=1}^M B_k$ dominates the dynamics. Its important to note that we do not make any other assumptions on a specific form of the observables $A$ and $B_k$. The resulting evolution is $U^{(M)}(t) = \sum_i \proj{i} \otimes \bigotimes_{k=1}^{M} U^{(k)}_{i}(t)$, where $U^{(k)}_{j}(t) \equiv \exp \left( - i a_j B_k t \right)$ and $\sum_{i=1}^{d_S} a_i \ket{i}\bra{i}\equiv A$. Discarding $(1-f)$-part of the environment we obtain a partially traced state:
\be
	\label{reduced} 
	\ba 
		& \varrho^{(fM)}(t) = \sum_{i=1}^{d_S} \sigma_{i} \proj{i} \otimes \bigotimes_{k=1}^{fM} \varrho^{(k)}_{i}(t) + \\ 
		& \sum_{i \neq j} \sigma_{ij} \prod_{k=1}^{(1-f)M} \gamma_{ij}^{(k)}(t) \kb{i}{j} \otimes \bigotimes_{k=1}^{fM} \varrho_{i,j}^{(k)}(t),
	\ea
\ee 
where $\varrho_{i,j}^{(k)}(t) \equiv U_{i}^{(k)}(t) \varrho^{(k)}(0) U_{j}^{(k) \dagger}(t)$, $\varrho_{i}^{(k)}(t) \equiv \varrho_{i,i}^{(k)}(t)$, $\sigma_{ij} \equiv \bra{i} \varrho(0) \ket{j}$, $\sigma_i \equiv \sigma_{i,i}$, and
$\gamma_{ij}^{(k)}(t) \equiv \Tr \left[ \varrho_{i,j}^{(k)}(t) \right]$ are the decoherence factors. We also define a collective decoherence factor: 
\be \label{G}
\Gamma(t) \equiv \sum_{i \neq j} \Ab{\sigma_{ij}} \prod_{k=1}^{(1-f)M} \Ab{\gamma_{ij}^{(k)}(t)}
\ee 
If the decoherence takes place, i.e. $\Gamma(t)= 0$, and for all $k=1,\dots,fM$ states $\varrho^{(k)}_{i}(t)$ and $\varrho^{(k)}_{j}(t)$ for $i \neq j$ become one-shot perfectly distinguishable, i.e. have orthogonal supports, we say that $\varrho^{(fM)}(t)$ has the SBS form. By measuring the supports of $\varrho^{(k)}_{i}(t)$ any of the local observers will then extract the same information about the state of the system, i.e. the index $i$, without disturbing (after forgetting the results) the state $\varrho^{(fM)}(t)$. Thus, with any SBS there is associated a natural local measurement for each environment $k$.

The main idea is to construct for each moment of time an SBS state which is naturally close to the above state \eqref{reduced}. Generally there is no obvious set of projectors the observers should use trying to discriminate the states $\varrho_{i}^{(k)}(t)$. Let us thus assume for a moment a generic situation where at the time $t$ the local observer monitoring the $k$-th environment uses some arbitrary, complete set of projectors $\mathsf P^{(k)}(t)\equiv \{ P_{i}^{(k)}(t)\}$. We will later optimize over these sets of projectors. If successful, $P_{i}^{(k)}(t)$ gives a state $P\varrho^{(k)}_{i}(t) \equiv P_{i}^{(k)}(t) \varrho^{(k)}_{i}(t) P_{i}^{(k)}(t)$, which happens with probability $p_{i}^{(k)}(t) = \Tr [ P\varrho^{(k)}_{i}(t) ]$. The probability that all the observers see the actual state $i$ is $r_{i}(t) \equiv \prod_{k=1}^{fM} p_{i}^{(k)}(t)$ and the cumulative error of discrimination over an ensemble $\{p_i,\varrho_{i}^{(k)}(t)\}$ reads \cite{Qiu}:
\be
	\label{eq:pError}
	p_{E} \left[ \left\{ p_{i},\varrho_{i}^{(k)}(t) \right\},\mathsf P^{(k)}(t) \right] \equiv \sum_{i} p_{i} \Tr \left[ \varrho_{i} \left( \openone - P_{i}^{(k)}(t) \right) \right].
\ee

Let us now construct a state that both attempts to approximate the actual state \eqref{reduced} and is SBS for the given set of local measurements $\mathsf P(t)\equiv\{\mathsf P^{(k)}(t)\}_{k=1}^{fM}$:
\be 
	\label{eq:broadcast}
	\varrho^{(fM)}_{SBS}(t;\mathsf P) \equiv \sum_{i} p_i(t) \proj{i} \otimes \bigotimes_{k=1}^{fM} \tilde{\varrho}^{(k)}_{i}(t), 
\ee 
where $p_i(t) \equiv \sigma_i r_i(t)/\sum_j \sigma_j r_j(t)$, cf. \eqref{reduced}, and $\tilde{\varrho}^{(k)}_i(t) \equiv P\varrho^{(k)}_{i}(t)/p_{i}^{(k)}(t)$. It is obtained by simply cutting the coherent part of \eqref{reduced} and projecting the environmental states $\varrho^{(k)}_{i}(t)$ on the subspaces defined by $P_i^{(k)}(t)$, thus making them perfectly distinguishable by the latter.

Our main result is an estimation of the distinguishability, given by the trace distance $\TN{\cdot}$, of the actual state \eqref{reduced} and the constructed ideal SBS \eqref{eq:broadcast}:

\begin{prop}
	\label{prop:main}
	For a given set of local measurements $\mathsf P(t)$, $\epsilon^{(fM)}(t;\mathsf P) \equiv \frac{1}{2} \TN{\varrho^{(fM)}(t) - \varrho^{(fM)}_{SBS}(t;\mathsf P)}$ we have
	\be
		\epsilon^{(fM)}(t;\mathsf P) \leq \Gamma(t) + \sum_{k=1}^{fM} p_{E} \left( \{ \sigma_{i}, \varrho^{(k)}_{i}(t) \}, \mathsf P^{(k)}(t) \right).
	\ee
\end{prop}

We present the proof in the Appendix~A. As one would expect, the lower the cumulative decoherence factor is and the more distinguishable are the environmental states $\varrho^{(k)}_{i}(t)$ by the projectors from $\mathsf P(t)$, the better the approximation of the state \eqref{reduced} by the constructed SBS \eqref{eq:broadcast} is. The above result quantifies this approximation. However, it depends on an arbitrary family of measurements used by the local observers and it would be desirable to obtain a more universal bound. To this end we use the result of of Barnum and Knill \cite{BarnumKnill,Montanaro}, providing an upper bound on the optimal discrimination error: 
\be
	\min_{\{ P_{i}\}} p_{E} \left( \{ p_{i},\varrho_{i}\}, \{ P_{i}\} \right) \leq \sum_{i \neq j} \sqrt{p_{i}p_{j}} B(\varrho_{i} ,\varrho_{j}),
\ee
where $B(\varrho,\sigma) \equiv \TN{ \sqrt{\varrho} \sqrt{\sigma} }$ is the fidelity \cite{Fuchs99}, which has already proven to be a convenient measure of the state distinguishability in the previous studies of SBS \cite{myPRL,TuziemskiKorbiczEPL,TuziemskiKorbiczPhotonics}. Optimizing over the discrimination measurements $\mathsf P(t)$ we obtain from the Proposition~\ref{prop:main}: 

\begin{cor}
	\label{cor1}
	The minimal distance of the actual state \eqref{reduced} to the SBS family \eqref{eq:broadcast} satisfies:
	\be
		\label{eq:cor1}
		\ba
			\frac{1}{2} & \min_{\mathsf P(t)}\TN{\varrho^{(fM)}(t) - \varrho^{(fM)}_{SBS}(t;\mathsf P)} \leq \eta \left[ \varrho^{(fM)}(t) \right] \\
			& \equiv \Gamma(t)+ \sum_{i \neq j} \sqrt{\sigma_i \sigma_j}\sum_{k=1}^{fM} B \left[ \varrho_{i}^{(k)}(t), \varrho_{j}^{(k)}(t) \right] .
		\ea
	\ee
\end{cor}

The essential power of the above construction is that it can be applied dynamically (i.e. for any fixed time $t$) in order to control how quickly the actual state can be approximated by some SBS state in the most fundamental sense, since the trace distance is related to the smallest error of global discriminating of them when we have the fifty-fifty ensemble $p_{E,\text{fifty-fifty}} = \frac{1}{2} \left(1 - \frac{1}{2} \TN{\varrho^{(fM)} - \varrho^{(fM)}_{SBS}} \right)$.

A second important corollary of the Proposition~\ref{prop:main} is a qualitative estimate on the information-theoretic condition for the objectivity of the quantum Darwinism \cite{Zurek10,ZurekNature,myPRL}. It relates the quantum mutual information, $I_{S:fM}$, between the system and the observed fraction of the environment, $I_{S:fM}[\varrho^{(fM)}(t)]$, to the von Neumann entropy, $S_{vN}$, of the decohered (pointer-diagonal) part of the system's state (cf. \eqref{reduced}), $S_{vN}[\varrho_{S,dec}]=H[\{\sigma_i\}]\equiv H_S$, where $H[\cdot ]$ is the Shannon entropy. Using the methods of \cite{myPRL} we prove in the Appendix~B the following:

\begin{cor}
	\label{cor2}
	Let $h(\epsilon)=-\epsilon\log\epsilon-(1-\epsilon)\log(1-\epsilon)$ be the binary Shannon entropy and $F(x) \equiv 4 h(2 x) + 2 h(x) + 10 x {{\log}_2 d_S}$. If for a given $t$ there exist a set of local measurements $\mathsf P$ such that $\epsilon^{(fM)}(t;\mathsf P) \leq 1/4$ then:
	\be
		\label{eq:cor2}
		\Ab{I_{S:fM}[\varrho^{(fM)}(t)]- H_{S}} \leq \min_{\mathsf P} F\left[\epsilon^{(fM)}(t;\mathsf P)\right].
	\ee
	If also $\eta \left[ \varrho^{(fM)}(t) \right] \leq 1/4$ then:
	\be
		\Ab{I_{S:fM}[\varrho^{(fM)}(t)]- H_{S}} \leq F\left[ \eta \left( \varrho^{(fM)}(t) \right) \right].
	\ee
\end{cor}

The latter statement in the Corollary~\ref{cor2} follows from the Corollary~\ref{cor1}. This result relates $\Ab{I[\varrho^{(fM)}(t)]- H_{S}}$ with decoherence and orthogonalization process efficiency.

\textit{\bf An example of objectivization process in the spin-spin model.-} We now show how the intersubjectivity of the state of the central system emerges in the spin-spin model, one of the canonical models of decoherence \cite{Zurek82,Zurek05,Zurek05b,SchlossauerBook}. The interaction Hamiltonian reads:
\be
	H_{int} = \frac{1}{2} \sigma_z \otimes \sum_{j=1}^N g_j \sigma_z^{(j)},
\ee
where $g_j$ are coupling constants and Pauli matrices $\sigma_z^{(j)}$ are acting on the space of the $j$-th spin. Decoherence in this model has been extensively studied in \cite{Zurek05}, assuming random, i.i.d. distributed couplings $g_j$, which we will also assume here. We divide the environmental spins into $M$ arbitrary fractions $mac_k$ \cite{myPRL} (it will become clear later why) and discard $(1-f)M$ of them as unobserved and show that there is a parameter regime that the joint state of the central spin~$S$ and the remaining $fM$ macrofractions has the SBS form i.e. $\varrho^{(fM)}(t) \approx \sum_{s = \pm} \sigma_{s} \proj{s} \otimes \bigotimes_{k=1}^{fM} \varrho^{mac_k}_s (t)$, with $\varrho^{mac_k}_{+} (t) \varrho^{mac_k}_{-} (t) \approx 0$, where $\{\ket{\pm}\}$ is the eigenbasis of $\sigma_z$, serving as pointer states \cite{Zurek81} and $\varrho_{\pm}^{mac}(t) = \otimes_{j \in mac} \rho^{(j)}_{\pm}(t)$, where $\rho^{(j)}_{\pm}(t)$ are states of the individual environmental spins.

\begin{figure}[ht]
	\begin{center}
		
		\includegraphics[scale=0.6]{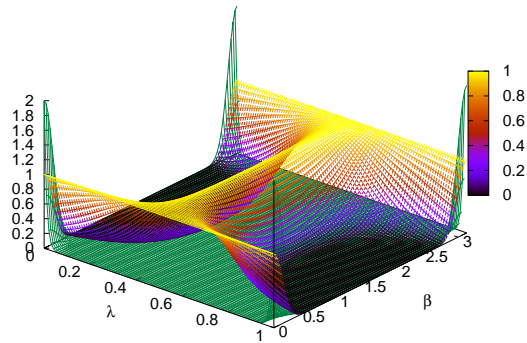}

		\caption{(Color on-line) Plot of $\E{B} \equiv \frac{1}{\tau} \int_0^\tau B(t) dt$ (colored surface), and $\E{\gamma} \equiv \frac{1}{\tau} \int_0^\tau \Ab{\gamma(t)} dt$ (green surface) as functions of the initial environmental state parameters $\lambda_+$ and $\beta$ for time $\tau$ large enough to dump the fluctuations. All the environmental spins are in the same initial state, $g_j$ are uniformly i.i.d. over $[0,1]$, and macrofractions have $100$ spins.
 \label{fig:3d}}
		
	\end{center}
\end{figure}

We use some of the methods of \cite{myPRL,TuziemskiKorbiczEPL}. Parametrizing the initial environmental states with Euler angles $\alpha_j, \beta_j, \gamma_j$, and an eigenvalue $\lambda_j$, we obtain that the only decoherence factor $\gamma (t)$ (cf. \eqref{reduced}) and the macrofraction state fidelity $B(t) \equiv B\left[\varrho_+^{mac}(t), \varrho_-^{mac}(t)\right]$ read:
\begin{subequations}
\label{eq:gammaB}
\be
		\label{eq:gamma2}
		\gamma(t) = \prod_{j=1}^{(1-f)N} \left[ \cos \left(g_j t\right) + i (2 \lambda_j - 1) \cos{\beta_j} \sin \left(g_j t\right) \right],
	\ee		
\be
		\label{eq:B}
		B(t) = \prod_{j \in mac} \sqrt{1 - (2 \lambda_j - 1)^2 \sin^2 \beta_j \sin^2(g_j t)}.
	\ee
\end{subequations}
It is now obvious that for an individual spin both functions are periodic with frequency $g_j$ and there is no chance for either decoherence or perfect state distinguishability. However, the coarse-graining of the environment into macrofractions together with random couplings turns the above functions into quasi-periodic ones. A sample plot of their long-time averages for macrofractions of $100$ spins is presented in Fig.~\ref{fig:3d}. Since both functions are positive, vanishing of the time averages is an indicator that the typical fluctuations of the functions above zero are small and hence an SBS appears \cite{myPRL}.


For large macrofractions it is possible to use the Law of Large Numbers \cite{TuziemskiKorbicz2016} and obtain the following analytical estimates (see Appendix~C): $B(t) \approx \exp \left[ -\Nm \overline{\kappa}(t) / 2 \right]$ and $\Ab{\gamma(t)}^2 \approx \exp \left[ - (1-f)N \overline{\chi}(t) \right]$, with $\Nm$ the macrofraction size and $\overline{\kappa}(t), \overline{\chi}(t) > 0$ for $t > 0$. The short-time behavior is $\overline{\kappa}(t) \approx (2/5) \overline{g^2} t^2$, and $\overline{\chi}(t) \approx (4/5) \overline{g^2} t^2$ with $\overline{g^2} \equiv \E{g^2_j}$ (we assume it exists and is i.i.d.), leading to the following timescales of decoherence $\tau_D = \sqrt{5 \log({\Nm})} / \sqrt{4 \overline{g^2}(1-f) N}$ and distinguishability $\tau_B = \sqrt{5 \log{(\Nm)}} / \sqrt{\overline{g^2} \Nm}$. The ratio $(\tau_B / \tau_D)^2 = 4 \frac{(1-f) N}{ \Nm}$ depends on the ratio of the traced out and the observed macrofractions.

\begin{figure}[t]
	\begin{center}
		
		\includegraphics[scale=0.6]{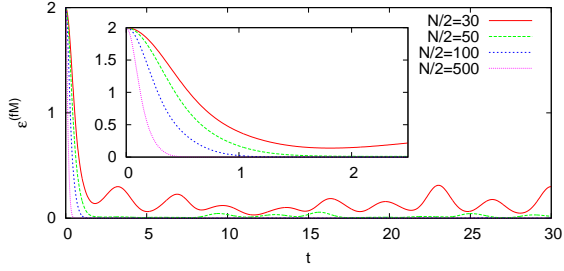}

		\caption{(Color on-line) The upper bound on $\frac{1}{2} \TN{\varrho^{(fM)}(t) - \varrho^{(fM)}_{SBS}(t)}$ obtained from the Corollary~\ref{cor1} using the optimal measurements and averaged over the Haar distribution of Euler angles, the uniform for the couplings and the Hilbert-Schmidt for the eigenvalues. We take equal size $N/2$ of the observed and unobserved macrofractions. In the pessimistic case the initial central state has $\sigma_{\pm} = \Ab{\sigma_{+-}} = 1/2$ yielding the bound $\gamma(t) + B(t)$ shown on the plot. For small macrofractions ($n=30,50$) the bound is substantially larger $1$ and does not imply an SBS formation. \label{fig:prop1}}
		
	\end{center}
\end{figure}

The formulas \eqref{eq:gammaB} can also be used to calculate the bound in Corollary~\ref{cor1}, we show an example for some cases in Fig.~\ref{fig:prop1}.

Let us now consider the SBS formation from the optimal measurement point of view, introduced in the first part of the paper. We need to show that there exist a measurement $\{P_{+}(t), P_{-}(t)\}$ which distinguishes with high probability between macrofraction states $\varrho_{\pm}^{mac}(t)$.

Take $P_{+}(t) \equiv \sum_{s \in S_{+}} \bigotimes_{j \in mac} P_{s_j}^{(j)}(t)$, where a set $S_{+} \subset \{+,-\}^{\Nm}$ with the majority of `$+$', and $P_{\pm}^{(j)}(t)$ are by our approach some projectors, which we will use to discriminate the states $\rho^{(j)}_{\pm}(t)$. We take $P_{-}(t) \equiv \openone - P_{+}(t)$ and call this a majority measurement, $\mathsf P^{maj}(t)$. For the local projectors $P_{\pm}^{(j)}(t)$ we use the construction of Helstrom \cite{Helstrom}: For $\varrho^{(j)}_{+}(t)$ and $\varrho^{(j)}_{-}(t)$, define $P^{(j)}_{+}(t)$ as the projector on the positive eigenspace of $\varrho^{(j)}_{+}(t) - \varrho^{(j)}_{-}(t)$, and $P^{(j)}_{-}(t) \equiv \openone - P^{(j)}_{+}(t)$. Using the same initial state parametrization as in \eqref{eq:gammaB} we find that:
\be
	\label{eq:rhokt}
	\rho_{\pm}^{(j)}(t) =
	\begin{bmatrix}
		\pi^{(j)}(t) & e^{\pm 2 i s g_j t} \delta_j \\
		e^{\mp 2 i s g_j t} \delta_j^{*} & 1 - \pi^{(j)}(t)
	\end{bmatrix},
\ee
with $\delta_j \equiv (1/2) \sin(\beta_j) \left[ e^{-i \alpha_j} \lambda_j - e^{i \gamma_j} (1 - \lambda_j) \right]$, and $\pi^{(j)}(t) \equiv (1/2)\left[ 1 + (2 \lambda_j - 1) \cos(\beta_j) \right]$, and hence the success probability of local distinguishing between $\rho_{\pm}^{(j)}(t)$ is $p_{+}^{(j)}(t) =p_{-}^{(j)}(t)\equiv 1/2 + \Ab{\delta_j} \Ab{\sin(2 g_j t)}$ (see Appendix~D).

Let us assume the initial conditions are i.i.d. (w.r.t. $j$), with $dg$ being the distribution of the couplings, and $d\mu$ of the environmental states. The average probability $\E{p}(t) \equiv \int dg \int d\mu p_{\pm}^{(j)}(t) \equiv 1/2 + \overline{S}(t)$ does not depend on $j$ and $\pm$, and for non-discrete measures is strictly greater than zero. We calculate the success probability of the majority measurement, $\tilde{p}(t)$. Due to the binary nature and the assumed i.i.d. $\tilde{p}(t)$ is equal to the probability that the majority of Bernoulli trials succeeded, with the success probability in each trial given by $\E{p}(t)$. The Chernoff bound then gives \cite{Sinclair}:
\be
	\label{eq:tildeP}
	\tilde{p}(t) \geq 1 - \exp \left[ -\frac{1}{2} \Nm \overline{S}(t)^2 \right].
\ee

One may ask what limits the above result imposes on the distinguishability of the states of macrofractions in the sense of Kolmogorov? The Kolmogorov distance between two discrete probability distributions is defined as $K(\{p_0(\cdot)\},\{p_1(\cdot)\}) = \frac{1}{2} \sum_x \Ab{p_0(x)-p_1(x)}$.
The probability distribution obtained with $\mathsf{P}^{maj}(t)$ on the state $\varrho_{+}^{mac}(t)$ and $\varrho_{\pm}^{mac}(t)$ is $[\tilde{p}(t), 1 - \tilde{p}(t)]$ and $[1 - \tilde{p}(t), \tilde{p}(t)]$, respectively. The Kolmogorov distance between these distributions is $\Ab{2 \tilde{p} - 1} \geq 1 - 2 \exp \left[ -(1/2) \Nm \overline{S}(t)^2 \right]$. On the other hand one knows \cite{Fuchs95,Fuchs99} that $\max_{\mathsf{P}} K \left[ \varrho_{+}(\mathsf{P}), \varrho_{-}(\mathsf{P}) \right] \leq 1 - \frac{1}{2} B \left(\varrho_{+}, \varrho_{-}\right)^2$, where $\varrho_{\pm}(\mathsf{P})$ is the probability distribution obtained with the measurement $\mathsf{P}$ on the state $\varrho_{\pm}$.
We saw that $B(t)$ decreases exponentially with the size of the macrofraction. Thus one cannot get better distinguishability than a one improving exponentially in the size of the macrofraction with coefficient defined by the parameters of the system. In this sense the majority measurement can be considered as optimal.

\textit{\bf Conclusions. -} The recent progress in the understanding of the emergence of the objectivity leads to a description where the information is in statu nascendi and at the same time system-environment state looks closer and closed to the SBS. However so far it has been only analyzed in its long time limit where the final state of the system together with a part of the environment reached the ideal SBS. Here we provided a natural way of estimation of the distance of the state to that ideal structure in a given moment of time.

Namely, as the Proposition~\ref{prop:main} states, if only the system-environment interaction is of the quantum measurement type then the distance of the state to the SBS can be bounded in terms of the degree of the success of a specific measurement aiming to discriminate some well defined states of different parts of the environment. If the probability of an error of the discrimination by all of the observers of this data is low, then the state has to be close to the SBS form. 

As shown in the Corollary~\ref{cor2} the above leads to a general way of monitoring how far is the process from the ideal information theoretic description of the information spreading into the environment in terms of mutual information condition considered originally in \cite{Zurek05,Zurek05b,Zurek10} and derived for two dimensional central systems from SBS in \cite{myPRL}.

We have also given an explicit example of the process of objectivization in the spin-spin model. We have shown that a simple majority measurement is enough to get close to optimal efficiency. This shows that even a measurement apparatus that independently measures the particles from a bulk interacting with it can lead to the intersubjectivity. This may have important consequences, e.g. for the theory of evolution of senses \cite{eye}, showing that the early eyes could have been simply a loosely coupled set of rough spots.

\textit{Acknowledgments. -} The work was made possible through the support of grant from the John Templeton Foundation. The opinions expressed in this publication are those of the authors and do not necessarily reflect the views of the John Templeton Foundation. It was also supported by a National Science Centre (NCN) grant 2014/14/E/ST2/00020.

\end{document}


%
%
%
%

In this Appendix we provide proofs of two propositions stated in the main text, give explicit calculations in the analysis of the spin-spin model and derive the exponential form of the formulas for the orthogonalization and decoherence.

\section{Appendix A: Proof of the Proposition 1}

The Proposition~1 states that for a given set of local measurements $\mathsf P(t)$, $\epsilon^{(fM)}(t;\mathsf P) \equiv \frac{1}{2} \TN{\varrho^{(fM)}(t) - \varrho^{(fM)}_{SBS}(t;\mathsf P)}$ we have
\be
	\epsilon^{(fM)}(t;\mathsf P) \leq \Gamma(t) + \sum_{k=1}^{fM} p_{E} \left( \{ \sigma_{i}, \varrho^{(k)}_{i}(t) \}, \mathsf P^{(k)}(t) \right),
\ee
where $\varrho^{(fM)}(t)$ is the partially traced state, $\varrho^{(fM)}_{SBS}(t;\mathsf P)$ is the SBS state constructed with the set of local projectors $\mathsf P$,  and $\Gamma(t)$ is the collective decoherence factor, see the main text for precise definitions. Here we rewrite the constructed SBS state as:
\be 
	\label{eq:broadcast}
	\varrho^{(fM)}_{SBS}(t;\mathsf P) = \frac{1}{\eta(t)} \sum_{i} \sigma_i \proj{i} \otimes \bigotimes_{k=1}^{fM} P\varrho^{(k)}_{i}(t) \equiv \frac{1}{\eta(t)} \tilde{\varrho}^{(fM)}_{SBS}(t),
\ee
with $\eta(t) \equiv \sum_j \sigma_j r_j(t)$ being the normalization of the projected state $\tilde{\varrho}^{(fM)}_{SBS}(t)$.

For the sake of clarity we omit here the explicit time dependence of states. We start the proof of the Proposition~1 with estimating the distance:
	\be
		\label{eq:prop1proof}
		\ba
			& \TN{\varrho^{(fM)} - \tilde{\varrho}^{(fM)}_{SBS}} \leq \TN{\sum_{i=1}^{d_S} \sigma_{i} \proj{i} \otimes \left[ {\bigotimes_{k=1}^{fM} \varrho^{(k)}_{i}} - {\bigotimes_{k=1}^{fM} \left( P\varrho^{(k)}_{i} \right)} \right] } + \Gamma  \leq \sum_{i=1}^{d_S} \sigma_{i} \TN{ {\bigotimes_{k=1}^{fM} \varrho^{(k)}_{i}} - {\bigotimes_{k=1}^{fM} \left( P\varrho^{(k)}_{i} \right)} } + \Gamma \\
			& \leq \sum_{i=1}^{d_S} \sigma_{i} \sum_{j=1}^{fM} \left[ \left( \prod_{k=1}^{j-1} \TN{ P\varrho^{(k)}_{i} } \right) \times \TN{\varrho^{(j)}_{i} - P\varrho^{(j)}_{i}} \times \left( \prod_{k=j+1}^{fM} \TN{\varrho^{(k)}_{i}} \right) \right] + \Gamma \leq \sum_{i=1}^{d_S} \sum_{j=1}^{fM} \sigma_{i} \TN{\varrho^{(j)}_{i} - P\varrho^{(j)}_{i}} + \Gamma,
		\ea
	\ee
where we used the following ``telescopic" inequality $\TN{ \bigotimes_{k=1}^{n} A^{(k)} - \bigotimes_{k=1}^{n} B^{(k)} } \leq \sum_{j=1}^{n} \left( \prod_{k=1}^{j-1} \TN{A^{(k)}} \right) \times \TN{A^{(j)} - B^{(j)}} \times \left( \prod_{k=j+1}^{n} \TN{B^{(k)}} \right)$, which follows immediately form the inductive application of the elementary inequality $\TN{A^{(1)} \otimes A^{(2)} - B^{(1)} \otimes B^{(2)}} \leq \TN{A^{(1)}} \TN{A^{(2)} - B^{(2)}} + \TN{A^{(1)} - B^{(1)}} \TN{A^{(2)}}$.

Now, using $\TN{\varrho^{(j)}_{i} - P\varrho^{(j)}_{i}} = \Tr \left[ \varrho^{(j)}_{i} \left( \openone - P_{i}^{(j)} \right) \right]$, we get that the last expression in \eqref{eq:prop1proof} is equal to
\be
	\sum_{i=1}^{d_S} \sum_{j=1}^{fM} \sigma_{i} \Tr \left[ \varrho^{(j)}_{i} \left( \openone - P_{i}^{(j)} \right) \right] + \Gamma,
\ee
or, using the formula for $p_{E}$, to
\be
	\sum_{i=1}^{d_S} \sum_{j=1}^{fM} p_{E} \left( \left\{ \sigma_{i},\varrho^{(j)}_{i} \right\}, \left\{ P_{i}^{(j)} \right\} \right) + \Gamma.
\ee
Recall that $\tilde{\varrho}^{(fM)}_{SBS} = \eta \varrho^{(fM)}_{SBS}$ with $\eta \in [0,1]$. We get the first part of the thesis of the Proposition 1 with the simple implication $\TN{\varrho - \eta \varrho'} \leq \upsilon \Rightarrow \TN{\varrho - \varrho'} \leq 2 \upsilon \Rightarrow \frac{1}{2} \TN{\varrho - \varrho'} \leq \upsilon$ (for any $\eta \in [0,1]$), following from the triangle inequality.

\section{Appendix B: Proof of the Corollary 2}
The main reasoning goes as in the Appendix of the paper \cite{myObj}. Let us choose an arbitrary set of local projectors $\mathsf{P}$ to be used in the construction of the SBS state $\varrho_{SBS}^{(fM)}(t)$ out of the partially traced state $\varrho^{(fM)}(t)$. Let us also fix a moment of time $t$. We would like to find the approximation of the quantity $\Ab{I_{S:fE} \left( \varrho^{(fM)} \right) - H_{S}}$, where $H_{S}$ is defined below and in the main text.

Define $\varrho_{S}(t)$ as the state of the central system with all the environments traced. Note that this state has $\sigma_i$ on the $i$-th diagonal position, and (possibly) some off-diagonal terms. Let $\varrho_{S,dec}(t)$ be the decohered state of the central system. The constructed SBS state after tracing the whole environment has zero off-diagonal terms and $p_i(t) \equiv \sigma_i r_i(t)/\sum_j \sigma_j r_j(t)$ on the diagonal; we denote this state of the central system by $\varrho_{S,SBS}^{(fM)}(t)$. Below we omit the explicit dependence of $\mathsf{P}$ and $t$ to avoid cumbersome notation.

Now, let us define
\be
	\eta_{S} \equiv \TN{\varrho_{S}- \varrho_{S,SBS}^{(fM)}} \leq \TN{ \varrho^{(fM)} - \varrho_{SBS}^{(fM)} } = 2 \epsilon^{(fM)},
\ee
and
\be
	\eta_{S,dec} \equiv \TN{\varrho_{S,dec} - \varrho_{S,SBS}^{(fM)}} \leq \TN{\varrho_{S} - \varrho_{S,SBS}^{(fM)}} \leq  \TN{\varrho^{(fM)} - \varrho_{SBS}^{(fM)}} = 2 \epsilon^{(fM)}.
\ee
Let $h(\epsilon)=-\epsilon\log\epsilon-(1-\epsilon)\log(1-\epsilon)$ be the binary Shannon entropy, $S_{vN}(\varrho)$ the von~Neumann entropy, $S_{S|fM}(\varrho)$ the conditional von~Neumann entropy between the system and the fraction of the environments and $I_{S:fM}(\varrho) = S_{vN}(\varrho_S) - S_{S|fM}(\varrho)$ the quantum mutual information between the system and the observed fraction of the environment. We have $S_{vN}[\varrho_{S,dec}]=H[\{\sigma_i\}]\equiv H_S$, where $H[\cdot]$ is the Shannon entropy. For any SBS state, by the construction, $I_{S:fM} \left[\varrho_{SBS}^{(fM)}\right] = S_{vN} \left[\varrho_{S,SBS}^{(fM)}\right]$, since SBS states reliably broadcast the spectrum of the central system.

Let us consider two states $\rho_1$ and $\rho_2$ with $T = \frac{1}{2} \TN{\rho_1 - \rho_2}$.
The Audenauert-Fannes \cite{Fannes,Audenaert} inequality states
\be
	\Ab{S_{vN}(\rho_1)-S_{vN}(\rho_2)} \leq T \log_2 (d_S - 1) + h(T) \equiv f^{(1)}(x,d_S).
\ee
The Alicki-Fannes \cite{AlickiFannes} inequality says
\be
	\Ab{S_{S|fM}(\rho_1) - S_{S|fM}(\rho_2)} \leq 8 T \log_2 (d_S) + 4 h(2 T) \equiv f^{(2)}(x,d_S).
\ee

Using the above notation we perform the following calculation:
\be
	\label{eq:cor2proof}
	\ba
		& \Ab{ I_{S:fM} \left(\varrho^{(fM)}\right) - H_{S} } = \Ab{ I_{S:fM} \left(\varrho^{(fM)}\right) - S_{vN} \left(\varrho_{S,dec}\right) - \left[I_{S:fM} \left(\varrho_{SBS}^{(fM)}\right) - S_{vN} \left(\varrho_{S,SBS}^{(fM)}\right) \right] } = \\
		& \Ab{ \left[ S_{vN} \left(\varrho_{S}\right) - S_{vN} \left(\varrho_{S,SBS}^{(fM)}\right) \right] - \left[ S_{S|fM}\left(\varrho^{(fM)}\right) - S_{S|fM}\left(\varrho_{SBS}^{(fM)}\right) \right] + \left[ S_{vN} \left(\varrho_{S,SBS}^{(fM)}\right) - S_{vN} \left(\varrho_{S,dec}\right) \right] },
	\ea
\ee
which is upper bounded by the mentioned inequalities $f^{(1)} \left(\eta_{S},d_{S}\right) + f^{(2)} \left(\epsilon^{(fM)},d_{S}\right) + f^{(1)} \left(\eta_{S,dec},d_{S}\right)$. Note that $f^{(1)}(x,d_S)$ and $f^{(2)}(x,d_S)$ are monotonically increasing for $x \leq 1/4$ and $\eta_{S},\eta_{S,dec} \leq \epsilon^{(fM)}$. Thus, if $\epsilon^{(fM)} \leq 1/4$ we can use the following upper bound on \eqref{eq:cor2proof}:
\be
	 2 f^{(1)} \left(\epsilon^{(fM)},d_{S}\right) + f^{(2)} \left(\epsilon^{(fM)},d_{S}\right).
\ee
This concludes the first part of the proof, since the set $\mathsf{P}$ was arbitrary.

The second part is a trivial consequence of the fact that the function $F(x) \equiv 4 h(2 x) + 2 h(x) + 10 x {{\log}_2 d_S}$ is monotically increasing for $x \leq \frac{1}{4}$ and by the Corollary~1 we have
\be
	\min_{\mathsf P(t)} \epsilon^{(fM)}(t;\mathsf P) \leq \eta \left[ \varrho^{(fM)}(t) \right].
\ee

\section{Appendix C: Exponential form of orthogonalization and decoherence factors and time scales of decoherence}
Now, we consider the asymptotic formulas for $B(t)$ and $\gamma(t)$. The rate of the convergence of the formulas depend on a particular way of dividing the environment into macrofractions, but the phenomenon of both decoherence and orthogonalization always occurs in the considered model for $N$ large enough. Rewriting the equation for $B(t)$ we get
\be
	\label{eq:asymptB}
	B(t) = {\exp \left( -\frac{1}{2} \sum_{j \in mac} \kappa_j \right)} \approx {\exp \left( -\frac{\Nm}{2} \overline{\kappa} \right)},
\ee
where $\kappa_j \equiv -\log \left[ 1 - (2 \lambda_j - 1)^2 \sin^2 \beta_j \sin^2 (g_j t) \right]$.

We denote by $\E{\cdot}$ the average value of a random variable. Since we assume that the parameters $\{\alpha_j, \beta_j, \gamma_j, \lambda_j, g_j\}$ are i.i.d. for different $j$, we have that $\overline{\kappa} \equiv \E{\kappa_j}$ does not depend on $j$.

We assume that $N$ and $\Nm$ are large enough to apply the Law of Large Numbers. For $\lambda_j \in [0,1]$ we have $\kappa_j > 0$ if $\lambda_j \neq 1$, $\beta_j \neq n \pi$ and $g_j t \neq n \pi$. For non-discrete probability distributions this happens with probability $1$, and thus $\overline{\kappa} > 0$, so for $\Nm \rightarrow \infty$ we have $B(t) \rightarrow 0$, and orthogonalization takes place.

Similarly, we rewrite $\Ab{\gamma(t)}^2$ as
\be
	\label{eq:asymptGamma}
	\Ab{\gamma(t)}^2 = \exp \left( - \sum_{j = 1}^{(1-f)N} \chi_j \right) \approx \exp \left[ - (1-f)N \overline{\chi} \right],
\ee
with $\chi_j \equiv -\log \left[ 1 + \sin^2 (g_j t) \left(-1 + (2 \lambda_j - 1)^2 \cos^2 \beta_j \right) \right]$, and $\overline{\chi} \equiv \E{\chi_j}$. Again, since $\chi_j > 0$ almost surely, we have $\overline{\chi} > 0$ for $t > 0$, and thus decoherence occurs.

Let us consider the case\footnote{This assumption holds if the process of creation of SBS is rapid, which is true for environments large enough.} when $t \ll 1$ with all coupling constants of the order $1$, thus also $g t \ll 1$.
We make now an additional assumption that the parameters of spins of the environment, $\{\alpha_j, \beta_j, \gamma_j\}$, are independent from $\lambda_j$ and the couplings $g_j$.
Let $\{\alpha_j, \beta_j, \gamma_j\}$ be distributed with the Haar measure,
\be
	\int_0^{\pi} d\beta \frac{\sin(\beta)}{2} \int_0^{2 \pi} d\alpha \frac{1}{2 \pi} \int_0^{2 \pi} d\gamma \frac{1}{2 \pi},
\ee
and $\lambda_j$ with the Hilbert-Schmidt measure \cite{HSmeas}, $\int_0^1 d\lambda 3 (2 \lambda - 1)^2$. Let us assume $\overline{g^2} \equiv \E{g^2_j} > 0$ (this average does not depend on $j$ by the assumption of i.i.d.). 

With the Haar measure we have $\E{ \sin^2 \beta_j} = 2/3$, $\E{\cos^2 \beta_j} = \frac{1}{3}$, and for the Hilbert-Schmidt measure $\E{(2 \lambda_j - 1)^2} = \frac{3}{5}$. Using the expansion $\log(1+x) \approx x$ and $\sin^2(x) \approx x^2$ for small $x$, we get $\kappa_j \approx (2 \lambda_j - 1)^2 \sin^2 \beta_j g_j^2 t^2$, and so $\overline{\kappa} \approx (2/5) \overline{g^2} t^2$. Similarly $\chi_j \approx - \left( -1 + (2 \lambda_j - 1)^2 \cos^2 \beta_j \right) g_j^2 t^2$, giving $\overline{\chi} \approx (4/5) \overline{g^2} t^2$.

Now, we investigate the time scales which can be observed in the evolution given by the above formulas. Let us fix some small value of $\epsilon$. From \eqref{eq:asymptB} it follows that $B(t) \approx \epsilon$ if, and only if $\overline{\kappa} \approx -\frac{2}{\Nm} \log{\epsilon}$, so that $t_B \approx \sqrt{\frac{5 \log{\frac{1}{\epsilon}}}{\overline{g^2} \Nm}}$. Taking $\epsilon = \frac{1}{\Nm}$ we get the following orthogonalization, or broadcasting, time:
\be
	\label{eq:tB}
	t_B = \sqrt{\frac{5 \log{\Nm}}{\overline{g^2} \Nm}}.
\ee

Now, we analyze \eqref{eq:asymptGamma} to investigate the time after which decoherence occurs, i.e. $\Ab{\gamma(t_D)}^2 \approx \epsilon$. It is easy to see that it holds when $\overline{\chi} \approx \frac{\log{\frac{1}{\epsilon}}}{(1-f)N}$.  From this we get that the time of decoherence is of the order $\sqrt{\frac{5 \log{\frac{1}{\epsilon}}}{4 \overline{g^2} (1-f) N}}$, and taking $\epsilon = \frac{1}{\Nm}$ we get the following formula for the decoherence time
\be
	\label{eq:tD}
	 t_D = \sqrt{\frac{5 \log{\Nm}}{4 \overline{g^2} N}}.
\ee
We see that the ratio $(t_B/t_D)^2 = 4 (1-f) N / \Nm$ depends on the relative size of the macrofraction comparing to the total environment.

\section{Appendix D: Objectivization process for spin-spin model}

The joint state of the central spin and a particular, arbitrary chosen, macrofraction under influence of the Hamiltonian $H_{int}$, is given by $\sum_{s = \pm} \sigma_s \proj{s} \otimes \varrho_s^{mac}(t)$, where $\varrho_s^{mac}(t) = \bigotimes_{j\in mac} \varrho_{\pm}^{(j)}(t)$. We consider orthogonalization of a particular macrofraction (which ones exactly is irrelevant for our analysis), after an interaction time $t$, $\varrho_s^{mac}(t) \equiv \bigotimes_{j\in mac} \varrho_{s}^{(j)}(t)$, with evolution $U_{s}^{(j)}(t)\equiv e^{i s g_j t \sigma_z^{(j)}}$. We calculate $B^{(j)}(t) \equiv B\left[ \varrho_{+}^{(j)}(t), \varrho_{-}^{(j)}(t) \right]$. Recall that we parametrize the initial states using the Euler angles:
\be
	\label{eq:RDR}
	\varrho_{+}^{(j)}(0) = \varrho_{-}^{(j)}(0) = \varrho^{(j)}(0) = R^{(j)} D^{(j)} R^{(j)\dagger},
\ee
with $D^{(j)} \equiv \text{diag}(\lambda_j, 1 - \lambda_j)$, and
\be
	\ba
		\label{eq:R}
		R^{(j)} \equiv
		\begin{bmatrix}
			e^{-\tfrac{i}{2}(\alpha_j + \gamma_j)}\cos\frac{\beta_j}{2} & -e^{-\tfrac{i}{2}(\alpha_j - \gamma_j)}\sin\frac{\beta_j}{2} \\
			e^{\tfrac{i}{2}(\alpha_j - \gamma_j)}\sin\frac{\beta_j}{2} & e^{\tfrac{i}{2}(\alpha_j + \gamma_j)}\cos\frac{\beta_j}{2}
		\end{bmatrix}.
	\ea
\ee
Then $B^{(j)}(t) =  \Tr \sqrt{{\bf M}^{(j)}}$, where we pulled some of the unitaries out of the square roots and used the cyclic property of the trace to get
\be
	\label{eq:M}
	{\bf M}^{(j)} \equiv \sqrt{D^{(j)}} R^{(j)\dagger} U_{-}^{(j)2} R^{(j)} D ^{(j)} R^{(j)\dagger} U_{+}^{(j)2} R^{(j)} \sqrt{D^{(j)}}.
\ee

Since ${\bf M}^{(j)}$ is two-dimensional it is enough to calculate its trace and determinant in order to perform the trace of the square root. We get $\Tr{\bf M}^{(j)} = \left[ \lambda_j^2 + (1-\lambda_j)^2 \right] - \left[ \lambda_j-(1-\lambda_j) \right]^2 \sin^2 \beta_j \sin^2(g_j t)$, and $\text{det}\left[{\bf M}^{(j)}\right] = \lambda_j^2 (1-\lambda_j)^2$. Denoting the eigenvalues of ${\bf M}^{(j)}$ by $M_{\pm}^{(j)}$ we have the usual formula:
\be
	M_{\pm}^{(j)} = \frac{1}{2} \left[ \Tr{\bf M}^{(j)} \pm \sqrt{ \left(\Tr{\bf M}^{(j)} \right)^2 - 4\text{det}{\bf M}^{(j)}} \right].
\ee
We have $B^{(j)}(t) = \prod_{j \in mac} \sqrt{M_{+}^{(j)}(t)} + \sqrt{M_{-}^{(j)}(t)} = \prod_{j \in mac} \sqrt{\Tr{\bf M}^{(j)}(t) + 2 \sqrt{\text{det}{\bf M}^{(j)}(t)}}$, and thus we derive the formula for $B^{(j)}(t)$. The Helstrom measurement for discrimination is given by the following projector measurement $\{P_{+}^{(j)}, P_{-}^{(j)}\}$:
\be
	P_{\pm}^{(j)}(t) =
	\begin{bmatrix}
		\frac{1}{2} & \pm \sgn \left[ \sin(2 g_j t) \right] \frac{i \delta_j}{2 \Ab{\delta_j}} \\
		\mp \sgn \left[ \sin(2 g_j t) \right] \times \frac{i \delta_j^{*}}{2 \Ab{\delta_j}} & \frac{1}{2}
	\end{bmatrix},
\ee
where $\delta_j \equiv (1/2) \sin(\beta_j) \left[ e^{-i \alpha_j} \lambda_j - e^{i \gamma_j} (1 - \lambda_j) \right]$. Now, it is easy to obtain that:
\be
	\label{eq:Pkt}
	p_{\pm}^{(j)} \equiv \Tr \left[ P_{\pm}^{(j)}(t) \varrho_{\pm}^{(j)}(t) \right] = \frac{1}{2} + \Ab{\delta_j} \times \Ab{\sin(2 g_j t)}.
\ee